\documentclass[aps,showpacs,preprintnumbers,amsmath, amssymb]{revtex4}

\usepackage{float}
\usepackage{graphics,epsfig}
\usepackage{graphicx}
\usepackage{dcolumn}
\usepackage{bm}

\begin{document}
\baselineskip=0.8 cm
\title{{\bf Notes on analytical study of holographic superconductors with Lifshitz scaling in external magnetic field}}

\author{Zixu Zhao$^{1,2}$, Qiyuan Pan$^{1,2}$\footnote{panqiyuan@126.com} and Jiliang Jing$^{1,2}$\footnote{jljing@hunnu.edu.cn}}
\affiliation{$^{1}$Institute of Physics and Department of Physics,
Hunan Normal University, Changsha, Hunan 410081, China}
\affiliation{$^{2}$ Key Laboratory of Low Dimensional Quantum
Structures and Quantum Control of Ministry of Education, Hunan
Normal University, Changsha, Hunan 410081, China}

\vspace*{0.2cm}
\begin{abstract}
\baselineskip=0.6 cm
\begin{center}
{\bf Abstract}
\end{center}

We employ the matching method to analytically investigate the
holographic superconductors with Lifshitz scaling in an external
magnetic field. We discuss systematically the restricted conditions
for the matching method and find that this analytic method is not
always powerful to explore the effect of external magnetic field on
the holographic superconductors unless the matching point is chosen
in an appropriate range and the dynamical exponent $z$ satisfies the
relation $z=d-1$ or $z=d-2$. From the analytic treatment, we observe
that Lifshitz scaling can hinder the condensation to be formed,
which can be used to back up the numerical results. Moreover, we
study the effect of Lifshitz scaling on the upper critical magnetic
field and reproduce the well-known relation obtained from
Ginzburg-Landau theory.

\end{abstract}

%\keywords{AdS/CFT correspondence, Black Holes in String Theory, Holography and condensed matter physics (AdS/CMT)}

\pacs{11.25.Tq, 04.70.Bw, 74.20.-z}\maketitle
\newpage
\vspace*{0.2cm}

\section{Introduction}
As the most successful realization of the holographic principle,
Maldacena first proposed the Anti-de Sitter/Conformal Field Theory
(AdS/CFT) correspondence \cite{Maldacena}, which has been a powerful
tool to deal with strongly coupled systems. In recent years, the
AdS/CFT correspondence has been applied to study the condensed
matter physics in order to understand the physics of high $T_{c}$
superconductors from the gravitational dual. Gubser first suggested
that black hole horizons could exhibit spontaneous breaking of an
Abelian gauge symmetry if the gravity was coupled to an appropriate
matter lagrangian, including a charged scalar that condenses near
the horizon \cite{Gubser}. Then Hartnoll, Herzog and Horowitz built
the first holographic superconductor model and reproduced the
properties of a ($2+1$)-dimensional superconductor in this simple
model \cite{HartnollPRL101}. The pioneering work on this topic has
led to many investigations concerning the condensation in bulk AdS
spacetime, for reviews, see Refs.
\cite{HerzogRev,HartnollRev,HorowitzRev} and references therein.

From the AdS/CFT correspondence, the AdS black hole geometry
corresponds to a relativistic CFT at finite temperature. However,
many condensed matter systems do not have relativistic symmetry.
Thus, Bu used the nonrelativistic AdS/CFT correspondence to study
the holographic superconductors in the Lifshitz black hole geometry
for $z=2$ in order to explore the effects of the dynamical exponent
and distinguish some universal properties of holographic
superconductors \cite{Bu}. It is found that the Lifshitz black hole
geometry results in different asymptotic behaviors of temporal and
spatial components of gauge fields than those in the
Schwarzschild-AdS black hole, which brings some new features of
holographic superconductor models. More recently, Lu $\emph{et al}.$
discussed the effects of the Lifshitz dynamical exponent $z$ on
holographic superconductors and gave some different results from the
Schwarzschild-AdS background \cite{Lu}. To this day, there have
attracted considerable interest to generalize the holographic
superconducting models to nonrelativistic situations
\cite{CaiLF,JingLF,Sin,Brynjolfsson,MomeniLifshitz,Schaposnik,Abdalla,Tallarita}.

On the other hand, according to the Ginzburg-Landau theory, it
should be noted that the upper critical magnetic field has the
well-known relation $B_{c}\propto(1-T/T_c)$ \cite{Poole}. Using the
semi-analytic method, Ge \emph{et al.} reproduced this relation in
the holographic superconductor model \cite{Ge}. As an important step
towards a realistic implementation of superconductivity through
holography, Dom\`{e}nech \emph{et al.} discussed the critical
magnetic fields in the holographic superconductor and analyzed the
effect of the dynamical magnetic field on the critical magnetic
field \cite{Salvio}. Cai \emph{et al.} studied the magnetic field
effect on the holographic insulator/superconductor phase transition
and found that the presence of the magnetic field causes the phase
transition hard \cite{CaiPRD}. Along this line, a number of attempts
have been made in order to investigate the effects of applying an
external magnetic field to holographic dual models
\cite{SetareEPL,GeBackreac,Montull,Gao,Ged,Roychowdhury,Momeni,Roychowdhury2,Cui,RoychowdhuryJHEP2013,Cai2013,Gangopadhyay}.
All these papers to study the effect of external magnetic field on
holographic dual models are made in relativistic situations. It is
therefore very natural to consider the nonrelativistic situations,
such as Lifshitz black hole.

In this work, we will use the matching method, which was first
proposed in \cite{Gregory} and later refined in \cite{PanWang}, to
analytically investigate the effect of external magnetic field on
holographic superconductors with Lifshitz scaling. We want to know
whether the relation $B_{c}\varpropto (1-T/T_c)$ can be reproduced
in holographic superconductor with Lifshitz scaling, and discuss the
effect of Lifshitz dynamical exponent $z$ on critical temperature as
well as the upper critical magnetic field. Furthermore, it is of
interest to analyze the restricted conditions for the matching
method since this issue has not been discussed systematically, and
examine whether the matching method is still valid to explore the
effect of the external magnetic field on the holographic
superconductor since we consider the totally different
nonrelativistic situations. We will concentrate on the probe limit
to avoid the complex computation in order to extract the main
physics.

The organization of the work is as follows. In Sec. II, we will
review the asymptotic Lifshitz black holes and study the holographic
superconductors with Lifshitz scaling. In Sec. III we investigate
the properties of the holographic superconductors with Lifshitz
scaling in an external magnetic field. We will conclude in the last
section of our main results.

\section{holographic superconductor models with Lifshitz scaling}

In order to study holographic superconductor with Lifshitz scaling,
we will present background for the gravity dual of the Lifshitz
fixed point. It is well known that there exist field theories with
anisotropic scaling symmetry between the temporal and spatial
coordinates $ t\rightarrow \lambda^z t,\quad x^i\rightarrow \lambda
x^i, $ which can be found in some condensed matter systems near the
critical point. From the generalized gauge/gravity correspondence,
one can attain scaling symmetry. The metric can be written as $
ds^2=L^2\left(-r^{2z}dt^2+r^2\sum_{i=1}^{d}dx_i^2+dr^2/r^2\right), $
where $0<r<\infty$ and $L$ is the radius of curvature of the
geometry. Kachru $\emph{et al}.$ first proposed this geometry
\cite{Kachru}, in which the action sourcing this geometry was also
given. The scale transformation is as follows
\begin{equation}
t\rightarrow \lambda^z t,\quad x^i\rightarrow \lambda x^i,\quad
r\rightarrow \frac{r}{\lambda},
\end{equation}
where $z$ is called the dynamical exponent. When $z=1$, the above geometry reduces to the usual $AdS_{d+2}$ spacetime.

The Lifshitz black holes can be constructed as in Ref.
\cite{Bertoldi} via the action
\begin{equation}
S=\frac{1}{16 \pi G_{d+2}}\int
d^{d+2}x\sqrt{-g}\left(R-2\Lambda-\frac{1}{2}\partial_{\mu}\Phi\partial^{\mu}
\Phi- \frac{1}{4}e^{\lambda\Phi}\mathcal {F}_{\mu\nu}\mathcal
{F}^{\mu\nu}\right),
\end{equation}
where $\Lambda=-(z+d-1)(z+d)/(2L^2)$ is the cosmological constant,
$\Phi$ is a massless scalar and $\mathcal{F}_{\mu\nu}$ is an abelian
gauge field strength. The geometry background is as follows
\begin{equation}
ds^2=L^2\left[-r^{2z}f(r)dt^2+r^2\sum_{i=1}^{d}dx_i^2+\frac{dr^2}{r^2f(r)}
\right],
\end{equation}
with $f(r)=1-r_+^{z+d}/r^{z+d}$. For
convenience, we set $L=1$ in the following discussion. Therefore the
Hawking temperature of the black hole is given by
\begin{equation}
T=\frac{z+d}{4\pi}r_+^z,\label{T}
\end{equation}
where $r_+$ is the radius of the event horizon.

In the probe limit, we will consider a Maxwell field and a charged
complex scalar field coupled via the action
\begin{eqnarray}\label{SWAction}
S=\int{d^{d+2}x\sqrt{-g}\left(-\frac{1}{4}F_{\mu\nu}F^{\mu\nu}
-|\nabla\psi-iA\psi|^{2}-m^2|\psi|^2\right)}.
\end{eqnarray}
Taking the ansatz of the matter fields as $\psi=\psi(r)$ and
$A=\phi(r) dt$, we can obtain the equations of motion from the
action (\ref{SWAction}) for the scalar field $\psi$ and gauge field
$\phi$
\begin{eqnarray}
 \psi''(r)+\left[\frac{f'(r)}{f(r)}+\frac{d+z+1}{r}\right]\psi'(r)+\left[\frac{\phi(r)^2}{r^{2z+2}f(r)^2}-
 \frac{m^2}{r^2f(r)}\right]\psi(r)=0,
\end{eqnarray}
\begin{eqnarray}
 \phi''(r)+\frac{d-z+1}{r}\phi'(r)-\frac{2\psi(r)^2}{r^2f(r)}\phi(r)=0.
\end{eqnarray}
Introducing $u=r_{+}/r$, we therefore have
\begin{eqnarray}
\psi''(u)+\left[\frac{f'(u)}{f(u)}+\frac{1-d-z}{u}\right]\psi'(u)+
\left[\frac{u^{2z-2} \phi(u)^2}{r_+^{2
z}f(u)^2}-\frac{m^2}{u^2f(u)}\right]\psi(u)=0,\label{Psieomz}
\end{eqnarray}
\begin{eqnarray}
\phi''(u)+\frac{z-d+1}{u}\phi'(u)-\frac{2\psi(u)^2
}{u^2f(u)}\phi(u)=0.\label{Phieomz}
\end{eqnarray}

At the horizon $u=1$, the regularity gives the boundary conditions
\begin{eqnarray}
\psi'(1)=-\frac{m^2}{z+d}\psi(1),~~~~~~\phi(1)=0.\label{Solutionz1}
\end{eqnarray}
Near the boundary ($u\rightarrow0$), the solutions behave as
\begin{eqnarray}
\psi(u)= J_{-}u^{\Delta_{-}} + J_{+}u^{\Delta_{+}},~~~~\phi(u) = \mu
- \rho\left(\frac{u}{r_+}\right)^{d-z},~(1\leq
z<d),\label{Solutionz0}
\end{eqnarray}
where the scaling dimension $\Delta_{\pm}$ of the scalar operator
dual to the bulk scalar $\psi$ is given by
$\Delta_{\pm}=[(z+d)\pm\sqrt{(z+d)^2+4m^2}]/2$, $ \mu $ and $ \rho $
are interpreted as the chemical potential and the charge density in
the dual field theory respectively. We impose boundary condition $
J_{-}=0 $ in the following discussion. For clarity, we set $
J=J_{+}$ and $ \Delta=\Delta_{+}$ in this work.

It should be noted that $\phi(u)=\rho-\mu\log u$ for the case $z=d$.
For simplicity, we will not consider this case in the analytical
studies, just as in Ref. \cite{Lu}.

Near $u=1$, the solution of $\psi(u) $ and $ \phi(u)  $ behave as
\begin{eqnarray}
\psi(u)=\psi(1)-\psi'(1)(1-u)+\frac{1}{2} \psi''(1)(1-u)^{2} +
....\label{Taylorpsi}
\end{eqnarray}
\begin{eqnarray}
\phi (u)=\phi (1) - \phi'(1)(1-u) + \frac{1}{2}\phi''(1)(1-u)^{2} +
.. ..\label{Taylorphi}
\end{eqnarray}

From (\ref{Psieomz}), (\ref{Phieomz}) and (\ref{Solutionz1}), we
obtain
\begin{equation}
\psi''(1)=\frac{m^2}{z+d}\left[1+\frac{m^2}{2(z+d)}\right]\psi(1)
-\frac{\phi'(1)^{2}}{2r_+^{2z}(z+d)^2}\psi(1)\label{PsiSecDer},
\end{equation}
\begin{equation}
\phi''(1)= -\phi'(1)\left[(z-d+1)+\frac{2\psi(1)^2}{z+d}\right].
\label{PhiSecDer}
\end{equation}

Substituting (\ref{Solutionz1}), (\ref{PsiSecDer}) and
(\ref{PhiSecDer}) into (\ref{Taylorpsi}) and (\ref{Taylorphi})
respectively, we have
\begin{equation}
\psi (u) =
\left(1+\frac{m^2}{z+d}\right)\psi(1)-\frac{m^2}{z+d}\psi(1)u+\frac{1}{4(z+d)^2}\left[m^4+2(z+d)m^2-\frac{\phi'(1)^{2}}{r_+^{2z}}\right]\psi(1)(1-u)^{2}\label{NewPsi},
\end{equation}
\begin{equation}
\phi(u)=
-\phi'(1)(1-u)-\frac{1}{2}\left[(z-d+1)+\frac{2\psi(1)^2}{z+d}\right]\phi'(1)(1-u)^{2}\label{NewPhi}.
\end{equation}

We need the following conditions so as to match the asymptotic
solutions at some intermediate point $ u=u_m$
\begin{eqnarray}
 Ju_m^{\Delta} =\left(1+\frac{m^2}{z+d}\right)\psi(1)-\frac{m^2}{z+d}\psi(1)u_m+\frac{1}{4(z+d)^2}\left[m^4+2(z+d)m^2-\frac{\phi'(1)^{2}}{r_+^{2z}}\right]\psi(1)(1-u_m)^{2},\label{Psium}
\end{eqnarray}
\begin{equation}
J{\Delta}u_m^{\Delta-1}
=-\frac{m^2}{z+d}\psi(1)-\frac{1}{2(z+d)^2}\left[m^4+2(z+d)m^2-\frac{\phi'(1)^{2}}{r_+^{2z}}\right]\psi(1)(1-u_m),
\label{Psiprimeum}
\end{equation}
\begin{eqnarray}
\mu - \rho\left(\frac{u_m}{r_+}\right)^{d-z}
=-\phi'(1)(1-u_m)-\frac{1}{2}\left[(z-d+1)+\frac{2\psi(1)^2}{z+d}\right]\phi'(1)(1-u_m)^{2},
\label{Phium}
\end{eqnarray}
\begin{equation}
-{\rho}(d-z)\frac{1}{r_+^{d-z}}u_m^{d-z-1}
=\phi'(1)+\left[(z-d+1)+\frac{2\psi(1)^2}{z+d}\right]\phi'(1)(1-u_m).
\label{Phiprimeum}
\end{equation}

Using (\ref{Psium}) and (\ref{Psiprimeum}), we have
\begin{equation}
J = \frac{u_m^{1-\Delta } \left[m^2 (u_m-1)-2(z+d)\right]}{(z+d)
[(\Delta -2) u_m-\Delta ]}\psi(1) ,
\end{equation}
and
\begin{eqnarray}\label{PhiAlpha}
&\phi'(1)&=-r_+^z \alpha\nonumber \\
&&=-r_+^z
\sqrt{m^4+2(z+d)m^2\left(\frac{2-u_m}{1-u_m}\right)+\frac{2\Delta
(z+d)[m^2(u_m-1)-2(z+d)]}{(1-u_m)[(\Delta-2)u_m-\Delta]}}.
\end{eqnarray}
In order to avoid a breakdown of the matching method, i.e., to
ensure that $\phi'(1)$ is real, it is interesting to observe that,
for different masses of the scalar field, from (\ref{PhiAlpha}) the
matching point $u_m$ has a range
\begin{eqnarray}\label{Range}
\left\{
\begin{array}{rl}
0< u_m <1   \ , &  \quad {\rm for} \ -(3-\sqrt{5})(z+d)\leq
m^{2}\leq 0, \\ \\ u_{md}< u_m <1  \ , &  \quad {\rm for} \
-\frac{(z+d)^{2}}{4}\leq m^{2}<-(3-\sqrt{5})(z+d),
\end{array}\right.
\end{eqnarray}
where we have defined the divergent point
\begin{eqnarray}
u_{md}=\frac{\Delta[m^4+6m^2(z+d)+4(z+d)^2]}{m^2(\Delta-1)(m^2+4z+4d)-\sqrt{m^2[m^2(m^2+4z+4d)^2-8\Delta(\Delta-2)(z+d)^3]}}.
\end{eqnarray}
This shows that the matching point is not truly arbitrary except in
the case of $-(3-\sqrt{5})(z+d)\leq m^{2}\leq 0$, which is
reminiscent of that seen for the Gauss-Bonnet holographic
superconductors \cite{PanWang}. Thus, if the mass of the scalar
field satisfies the inequality $-\frac{(z+d)^{2}}{4}\leq
m^{2}<-(3-\sqrt{5})(z+d)$, the matching point has to be in an
appropriate range of values which depend on Lifshitz scaling,
spacetime dimension and scalar mass.

From (\ref{Phiprimeum}) we obtain
\begin{equation}
\psi(1)^{2}=
\frac{z+d}{2(1-u_m)}\left[-\frac{(d-z)u_m^{d-z-1}\rho}{\phi'(1)r_+^{d-z}}-(z-d+1)(1-u_m)-1\right]\label{Psi1Squ}.
\end{equation}
Considering (\ref{T}), from (\ref{Psi1Squ}) we have
\begin{equation}
\psi(1)^{2}=
\frac{(z+d)[1+(z-d+1)(1-u_m)]}{2(1-u_m)}\left(\frac{T_c}{T}\right)^{\frac{d}{z}}\left[1-\left(\frac{T}{T_c}\right)^{\frac{d}{z}}\right]\label{NewPsiSqu},
\end{equation}
where the critical temperature $T_c$ is given by
\begin{equation}
T_c=\frac{z+d}{4\pi}\left[\frac{(d-z)u_m^{d-z-1}\rho}{\alpha[1+(z-d+1)(1-u_m)]}\right]^{\frac{z}{d}}\label{Tcexp}.
\end{equation}
It should be noted that, besides the constraint condition
(\ref{Range}), in order to ensure the accuracy and correctness of
the calculations for $T_{c}$ we require another constraint
\begin{eqnarray}\label{RangeMid}
\frac{d-(z+2)}{d-(z+1)}< u_m <1,
\end{eqnarray}
which depends only on Lifshitz scaling $z$ and spacetime dimension
$d$. According to (\ref{NewPsiSqu}), the constraint (\ref{RangeMid})
can be used to ensure that $\psi(1)$ is real. Considering the
possibility $u_{md}<\frac{d-(z+2)}{d-(z+1)}$, for the matching point
$u_m$ we arrive at
\begin{eqnarray}\label{RangeTc}
\max\left[0,u_{md},\frac{d-(z+2)}{d-(z+1)}\right]< u_m <1,
\end{eqnarray}
which will lead to the correct critical temperature $T_{c}$. Thus,
the matching point $u_m$, which depends on Lifshitz scaling $z$,
spacetime dimension $d$ and scalar mass $m$, is not truly arbitrary
and must obey the constraint (\ref{RangeTc}). This means that, from
Eqs. (\ref{NewPsi}) and (\ref{NewPhi}), the asymptotic solutions
$\psi(u)$ and $\phi(u)$ both are physical solutions. For the case of
$z=1$ and $d=2$, it is to be noted that we can easily obtain
$0<u_{m}<1$ for all the scalar masses.

For concreteness, we choose $z=1$ and $z=2$ with $d=3$, $m^2=-3$ and
$u_m=1/2$ which satisfies the range given in (\ref{RangeTc}) and get
\begin{equation}
T_c(z=1)
=\frac{1}{\pi}\left(\frac{1}{2}\right)^{\frac{1}{3}}\left(\frac{5}{309}\right)^{\frac{1}{6}}{\rho}^{\frac{1}{3}}\approx
0.202{\rho}^{\frac{1}{3}},
\end{equation}
\begin{equation}
T_c(z=2)
=\frac{5}{4\pi}\left(\frac{79-20\sqrt{13}}{1041}\right)^{\frac{1}{3}}{\rho}^{\frac{2}{3}}\approx0.0747{\rho}^{\frac{2}{3}}.
\end{equation}
Obviously, $T_c(z=2)$ is smaller than $T_c(z=1)$, which means that
the larger dynamical exponent $z$ makes the condensation harder to
form. This tendency is the same as found in Ref. \cite{Lu}. For
$z=1,~d=2,~m^2=-2$ and $u_m=1/2$, it is to be noted that our result
reduces to $T_c=\frac{3\sqrt{\rho}}{4\pi\sqrt{2\sqrt{7}}}$, which is
obtained in Refs. \cite{Gregory,Ge}.

Following the AdS/CFT dictionary, near the critical temperature $
T\sim T_c $ we can express the relation for the condensation
operator $\langle\mathcal{O}\rangle=Jr^{\Delta}_{+}$ as
\begin{eqnarray}
\langle\mathcal{O}\rangle^{\frac{1}{\Delta}}&=&\left(\frac{4\pi
T_c}{z+d}\right)^{\frac{1}{z}} \left\{\frac{u_m^{1-\Delta }
\left[m^2 (u_m-1)-2(z+d)\right]}{(\Delta -2) u_m-\Delta
}\right\}^{\frac{1}{\Delta}}\left[\frac{1+(z-d+1)(1-u_m)}{2(z+d)(1-u_m)}\right]^{\frac{1}{2\Delta}}
\left[1-\left(\frac{T}{T_c}\right)^{\frac{d}{z}}\right]^{\frac{1}{2\Delta}}.\nonumber \\
\end{eqnarray}
The analytic result supports the numerical computation \cite{Lu}
that the phase transition of holographic superconductors with
Lifshitz scaling belongs to the second order and the critical
exponent of the system takes the mean-field value 1/2. The Lifshitz
scaling and spacetime dimension will not influence the result.

\begin{figure}[H]
\begin{centering}
\includegraphics[scale=0.7]{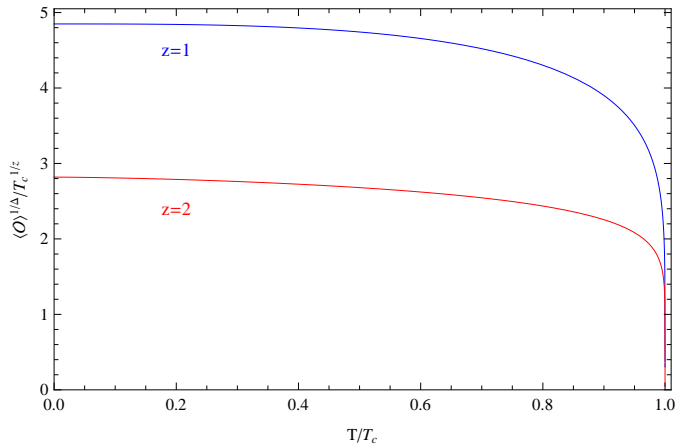}\hspace{0.2cm}%
\caption{\label{Tc} (color online) The condensate of the scalar
operator as a function of $T/T_c$ obtained by using the analytic
matching method. We choose the mass of the scalar field by
$m^{2}=-3$ for the fixed $d=3$ and $u_m=1/2$. The top line
corresponds to $z=1$ (blue) and bottom one is $z=2$ (red).}
\end{centering}
\end{figure}

Fixing $d=3$ and $u_m=1/2$, in Fig. \ref{Tc} we present the
condensate of the scalar operator $\langle\mathcal{O}\rangle$ as a
function of temperature with different dynamical exponent $z$ for
the mass of the scalar field $m^{2}=-3$. From Fig. \ref{Tc}, we see
that the gap becomes smaller as $z$ increases, which corresponds to
the lower critical temperature. This agrees with the numerical
results obtained in \cite{Lu}. It implies that the matching method
is still powerful to study the holographic superconductors in
Lifshitz black hole.

\section{Effect of external magnetic field on superconductor models with Lifshitz scaling}

Now we are in a position to study the effect of external magnetic
filed on the holographic superconductors with Lifshitz scaling. From
the gauge/gravity correspondence, the asymptotic value of the
magnetic field corresponds to a magnetic field added to the boundary
field theory. Near the upper critical magnetic field $B_c$, the
scalar field $\psi$ can be regarded as a perturbation. Following
\cite{Albash}, we set the ansatz
\begin{equation}
A=\phi(u)dt+Bxdy,~~~\psi=\psi(x,u).
\end{equation}
We therefore obtain the scalar field equation for $\psi$
\begin{eqnarray}
&&\psi''(x,u)+\left[\frac{f'(u)}{f(u)}+\frac{1-d-z}{u}\right]\psi'(x,u)\nonumber \\
&&+ \left[\frac{{u^{2z-2}\phi(u)}^2}{r_+^{2z}f(u)^2
}-\frac{m^2}{u^2f(u)}\right]\psi(x,u)+\frac{1}{r_+^2f(u)}(\partial_{x}^{2}
- B^{2}x^{2})\psi(x,u)=0\label{PsiB}.
\end{eqnarray}

Eq. (\ref{PsiB}) can be solved by separating the variables separable
form
\begin{equation}
\psi (x,u)=X(x)R(u)\label{XR}.
\end{equation}
Substituting (\ref{XR}) into (\ref{PsiB}), we can get
\begin{eqnarray}
r_{+}^{2}f(u)\left\{\frac{R''(u)}{R(u)}+
\left[\frac{f'(u)}{f(u)}+\frac{1-d-z}{u}\right]\frac{R'(u)}{R(u)}+
\left[\frac{{u^{2z-2}\phi(u)}^2}{r_+^{2z}f(u)^2}-\frac{m^2}{u^2f(u)}\right]\right\}-\left[-\frac{X''(x)}{X(x)}+B^{2}x^{2}\right]=0.
\end{eqnarray}
The equation for $X(x)$ can be considered as the
Schr$\ddot{o}$dinger equation in one dimension with frequency
determined by $B$ \cite{Albash}
\begin{equation}
-X''(x) + B^{2}x^{2}X(x) = \lambda_{n} B X(x),
\end{equation}
where $\lambda_{n}= 2n+1$ denotes the separation constant. We
consider the lowest mode ($ n=0 $) solution, which is the first to
condensate and the most stable solution after condensation
\cite{Albash}. Thus, we can express the equation of $ R(u) $ as
\begin{equation}
R''(u)+\left[\frac{f'(u)}{f(u)}+\frac{1-d-z}{u}\right]R'(u)+\left[\frac{u^{2z-2}{\phi(u)}^2
}{r_+^{2z}f(u)^2}-\frac{m^2}{u^2f(u)}-\frac{B
}{r_+^2f(u)}\right]R(u)=0\label{REOM}.
\end{equation}

At the horizon ($ u=1 $), from Eq. (\ref{REOM}), we have
 \begin{equation}
 R'(1) =-\frac{1}{z+d}\left(m^2+\frac{B}{r_+^2}\right)R(1)\label{Rd1}.
 \end{equation}
The asymptotic behavior ($ u\rightarrow 0 $) for (\ref{REOM}) can be expressed as
 \begin{equation}
 R(u) = J_- u^{\Delta_-} + J_+ u^{\Delta_+} \label{R0}.
 \end{equation}
In the following calculation we still let $J_{-}=0$ and set
$J=J_{+}$ and $ \Delta=\Delta_{+}$ just as discussed in the previous
section.

Near the horizon $u=1$, we can expand $R(u)$ in a Taylor series as
\begin{equation}
R(u)=R (1) - R'(1)(1-u) + \frac{1}{2} R''(1)(1-u)^{2} + ..
..\label{R1}
 \end{equation}
From (\ref{REOM}), we have
\begin{equation}
R''(1) =
\frac{1}{(z+d)^2}\left[m^2\left(z+d+\frac{m^2}{2}\right)-\frac{\phi'(1)^{2}}{2
r_+^{2z}}
+\frac{Bm^2}{r_+^{2}}+\frac{B^{2}}{2r_+^{4}}\right]R(1)\label{Rd2}.
\end{equation}
Substituting  (\ref{Rd1}) and (\ref{Rd2}) into (\ref{R1}), we get
the approximate solution
 \begin{eqnarray}
R(u)&=&R(1)+\frac{1}{z+d}\left(m^2+\frac{B}{r_+^2}\right)R(1)(1-u)\nonumber \\
&&+\frac{1}{2(z+d)^2}\left[m^2\left(z+d+\frac{m^2}{2}\right)-\frac{\phi'(1)^{2}}{2
r_+^{2z}}
+\frac{Bm^2}{r_+^{2}}+\frac{B^{2}}{2r_+^{4}}\right]R(1)(1-u)^{2}\label{Ru}.
 \end{eqnarray}

Matching (\ref{R0}) and (\ref{Ru}) for $J_{-}=0$ at some
intermediate point $u=u_m$, we have the following two equations
\begin{eqnarray}
Ju_m^{\Delta}&=&\left[1+\frac{1}{z+d}\left(m^2+\frac{B}{r_+^2}\right)\right]R(1)- \frac{1}{z+d}\left(m^2+\frac{B}{r_+^2}\right)R(1)u_m\nonumber \\
&&+
\frac{1}{2(z+d)^2}\left[m^2\left(z+d+\frac{m^2}{2}\right)-\frac{\phi'(1)^{2}}{2
r_+^{2z}}
+\frac{Bm^2}{r_+^{2}}+\frac{B^{2}}{2r_+^{4}}\right]R(1)(1-u_m)^{2},
\label{NewMaRz0}
\end{eqnarray}
\begin{equation}
J{\Delta}u_m^{{\Delta}-1} =
-\frac{1}{z+d}\left(m^2+\frac{B}{r_+^2}\right)R(1)-\frac{1}{(z+d)^2}\left[m^2\left(z+d+\frac{m^2}{2}\right)-\frac{\phi'(1)^{2}}{2
r_+^{2z}}
+\frac{Bm^2}{r_+^{2}}+\frac{B^{2}}{2r_+^{4}}\right]R(1)(1-u_m)\label{NewMaRPrz0},
\end{equation}
which give a solution
\begin{eqnarray}
B=\frac{r_+^{2}}{(u_m-1) [(\Delta -2) u_m-\Delta
]}\left\{\sqrt{\gamma+(u_m-1)^2 [\Delta -(\Delta -2)
u_m]^2\left[\frac{\phi'(1)}{r_+^{z}}\right]^2}-\beta \right\},
\label{Bc2}
\end{eqnarray}
with
\begin{eqnarray}
&&\beta=2 u_m \left[(z+d)-m^2(u_m-1)\right]-\Delta  (u_m-1) \left[2(z+d)-m^2(u_m-1)\right], \nonumber \\
&&\gamma=2(z+d)\left\{2u_m^2(z+d)-m^2(u_m-1)^2[\Delta
-(\Delta-2)u_m]^2\right\}.
\end{eqnarray}

When the external magnetic field is very close to the upper critical
magnetic field $B_{c}$, the condensation is so small that we can
ignore all the quadratic terms in $\psi$ and Eq. (\ref{Phieomz})
reduces to
\begin{equation}
\phi''(u)+\frac{z-d+1}{u}\phi'(u)=0.
\end{equation}
We can obtain
\begin{equation}
\phi(u) = \mu -
\rho\left(\frac{u}{r_+}\right)^{d-z},
\end{equation}
which results in
\begin{equation}
\phi'(1) = -\frac{\rho}{{r_+}^{d-z}}(d-z)\label{PhiPrimez1}.
\end{equation}
Using (\ref{T}), (\ref{Tcexp}) and (\ref{PhiPrimez1}), we can
express the critical magnetic field $B_{c}$ as
\begin{eqnarray}
B_{c}&=&\left(\frac{4\pi
T}{z+d}\right)^{\frac{2}{z}}\left(\frac{T_{c}}{T}\right)^{\frac{d}{z}}\frac{1}{(u_m-1)
[(\Delta -2) u_m-\Delta
]} \nonumber \\
&&\times\left\{\sqrt{(\beta^{2}-\gamma)u_m^{2(1+z-d)}[1+(1+z-d)(1-u_m)]^{2}
+\gamma\left(\frac{T}{T_{c}}\right)^{\frac{2d}{z}}}-\beta\left(\frac{T}{T_{c}}\right)^{\frac{d}{z}}\right\}.
\label{BcTTc}
\end{eqnarray}

\begin{figure}[H]
\begin{centering}
\includegraphics[scale=0.41]{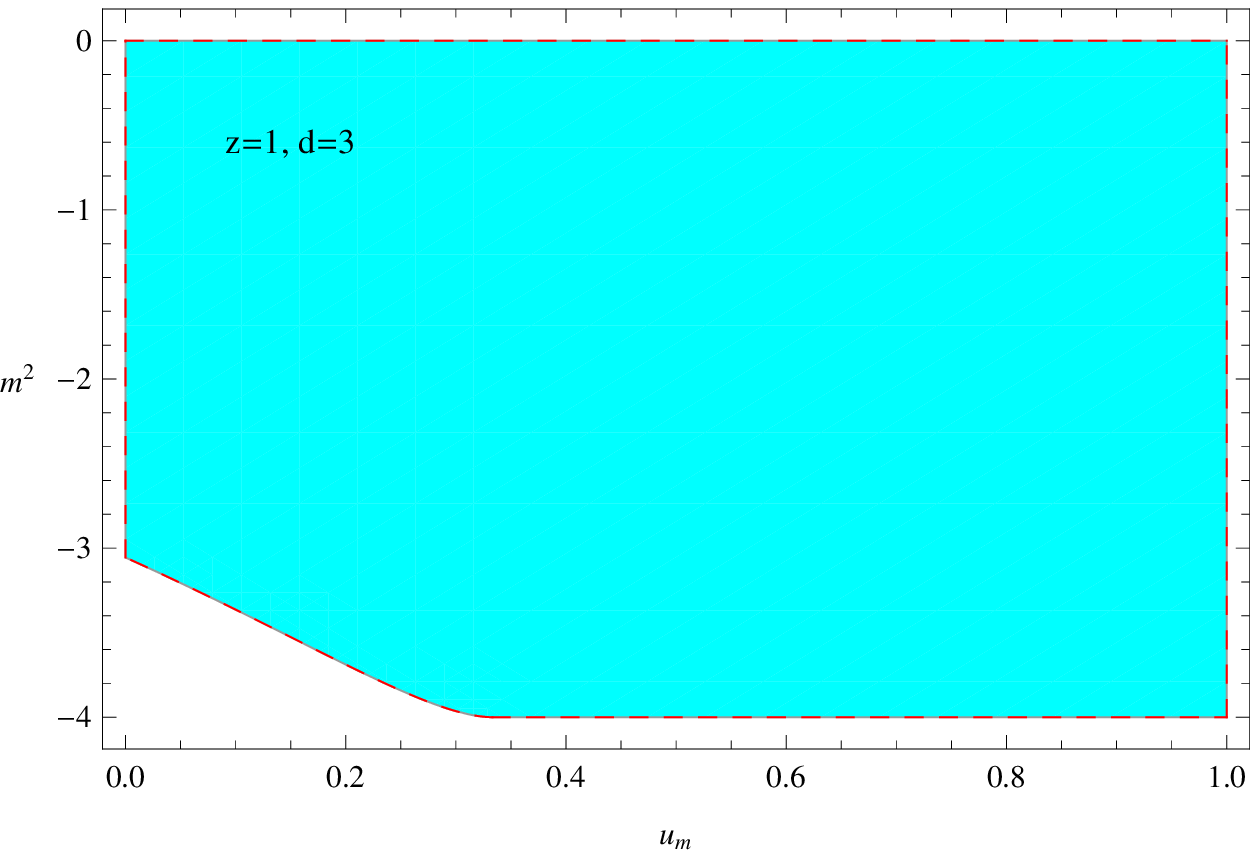}\vspace{0.0cm}
\includegraphics[scale=0.41]{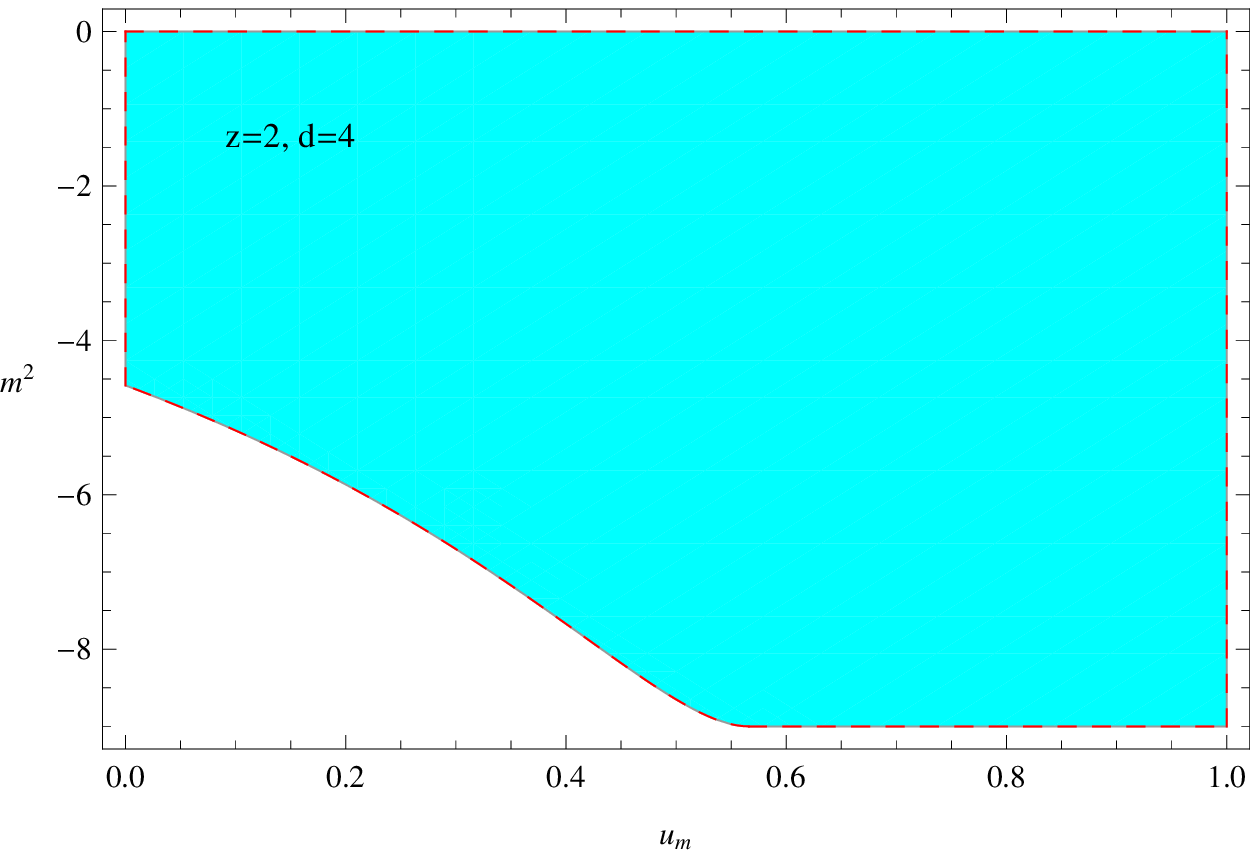}\vspace{0.0cm}
\includegraphics[scale=0.41]{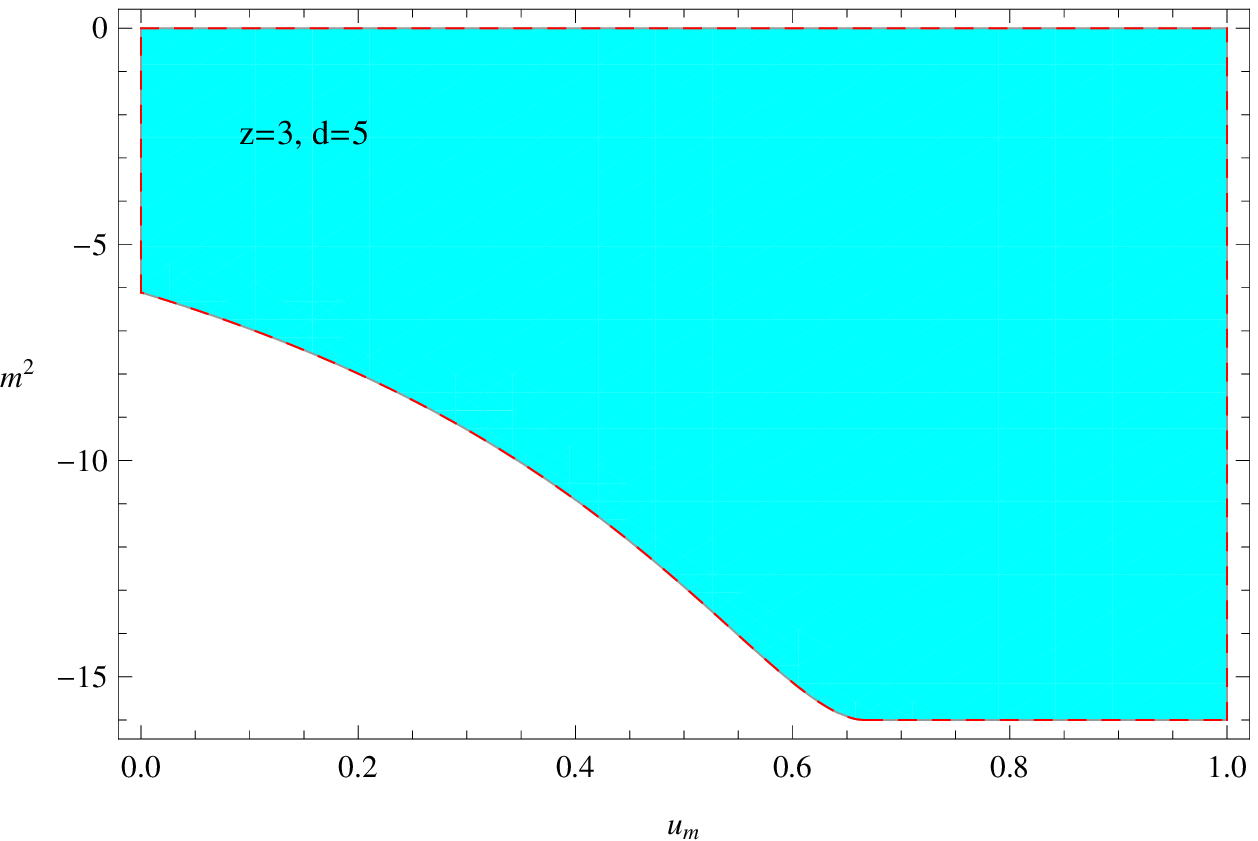}\\ \vspace{0.0cm}
\includegraphics[scale=0.41]{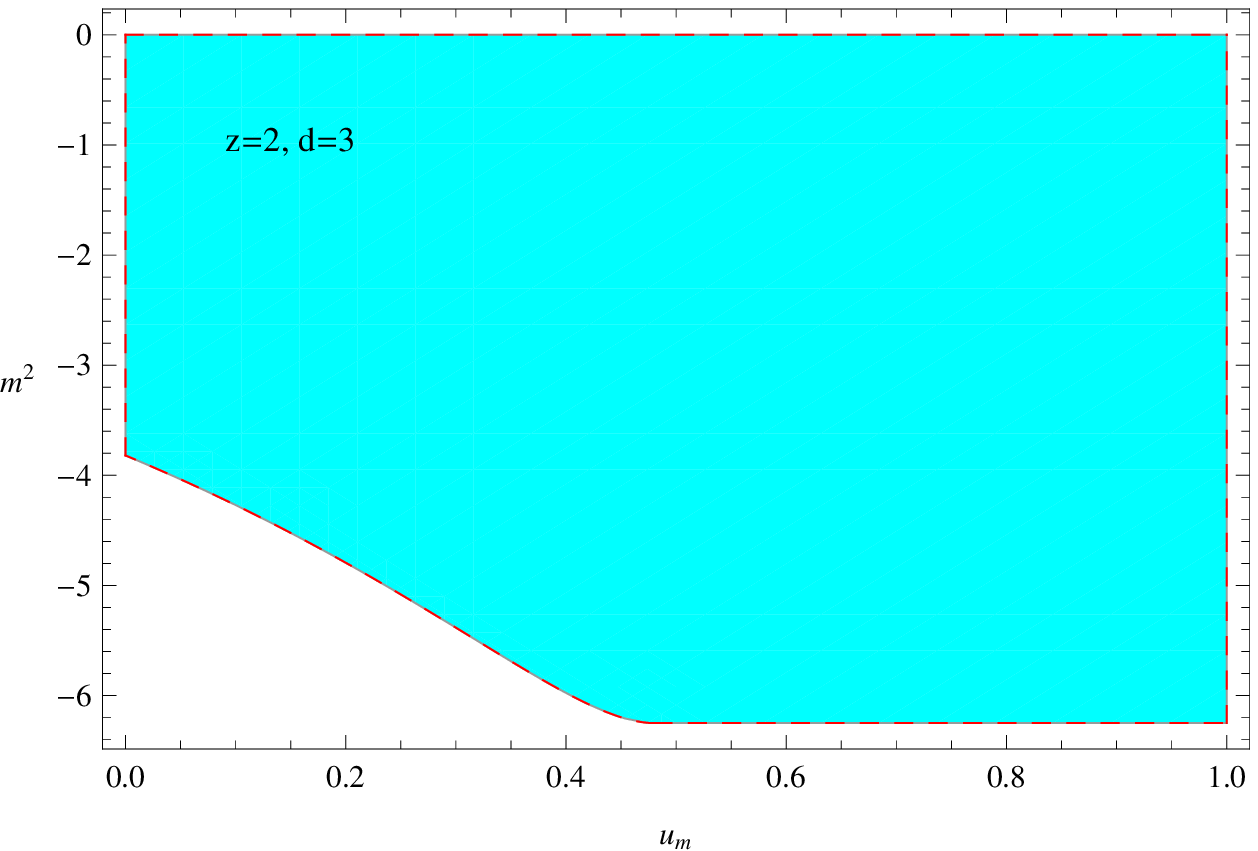}\vspace{0.0cm}
\includegraphics[scale=0.41]{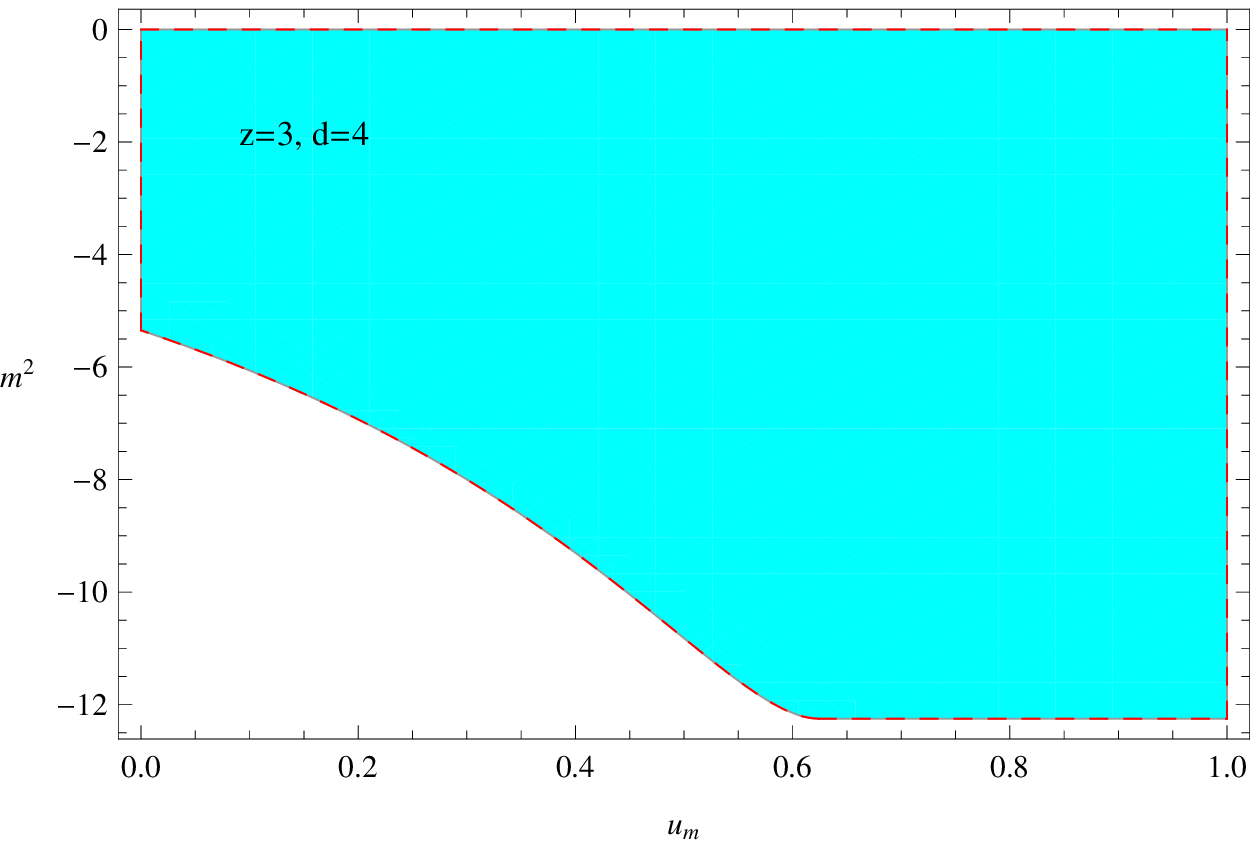}\vspace{0.0cm}
\includegraphics[scale=0.41]{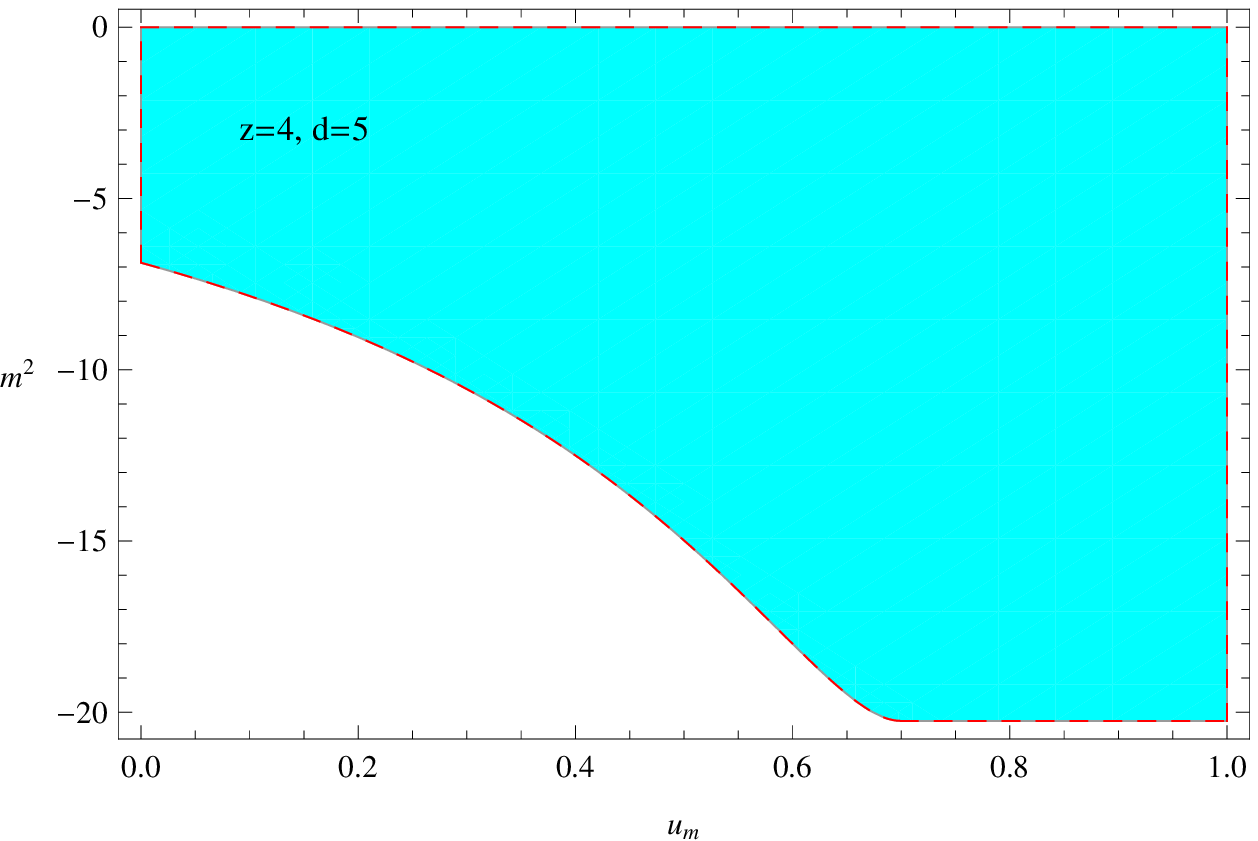}\\ \vspace{0.0cm}
\caption{\label{RegionA} (color online) The allowable range of $u_m$
for different masses if $z=d-1$ and $z=d-2$. In each panel, the
region surrounded by the red and dashed line is determined by Eq.
(\ref{Range}), and the cyan region corresponds to Eq.
(\ref{RangeTc}). In this case, the study of $B_{c}$ does not bring
new restriction on the selection of $u_m$.}
\end{centering}
\end{figure}

Note that there is a superconducting phase transition when $B_{c}=0$
at $T=T_{c}$, we have
\begin{eqnarray}
\gamma=\beta^{2}, \label{gammabeta}
\end{eqnarray}
which is related to Lifshitz scaling, spacetime dimension and the
scalar mass, or
\begin{eqnarray}
u_m^{2(1+z-d)}[1+(1+z-d)(1-u_m)]^{2}=1, \label{RelationBc}
\end{eqnarray}
which is related to Lifshitz scaling and spacetime dimension but
independent of the scalar mass. For Eq. (\ref{gammabeta}), the only
root which is probably in the range $0<u_m<1$ is $u_m=u_{md}$ where
the scalar mass satisfies $-(z+d)^{2}/4\leq
m^{2}<-(3-\sqrt{5})(z+d)$. But from Eq. (\ref{RangeTc}) we know that
this fixed matching point $u_{md}$ will cause a breakdown of the
matching method. So we have to count on Eq. (\ref{RelationBc})
instead of Eq. (\ref{gammabeta}) to ensure the condition $B_{c}=0$
at $T=T_{c}$. For Eq. (\ref{RelationBc}), it is interesting to find
that if
\begin{eqnarray}
z=d-1,~~~{\rm or}~~~z=d-2, \label{RelaZD}
\end{eqnarray}
the relation (\ref{RelationBc}) always holds for all $u_m$ selected
in the range (\ref{RangeTc}) which is shown in Fig. \ref{RegionA}.
That is to say, for the case (\ref{RelaZD}), in an appropriate range
(\ref{RangeTc}) we can choose the matching point $u_m$ arbitrarily.
It should be noted that in this case the investigation of the
critical magnetic field $B_{c}$ does not bring new restriction on
the selection of the matching point $u_m$. From Fig. \ref{RegionA},
we clearly find that the allowable range of the matching point $u_m$
depends on Lifshitz scaling $z$, spacetime dimension $d$ and scalar
mass $m$. For the Breitenlohner-Freedman (BF) bound
$m^{2}=-(z+d)^{2}/4$ \cite{BreitenloherFreedman}, it shows that the
range of $u_m$ becomes smaller as we amplify the value of $z$ for
the fixed $d$ or increase the value of $d$ for the fixed $z$.

\begin{figure}[H]
\begin{centering}
\includegraphics[scale=0.41]{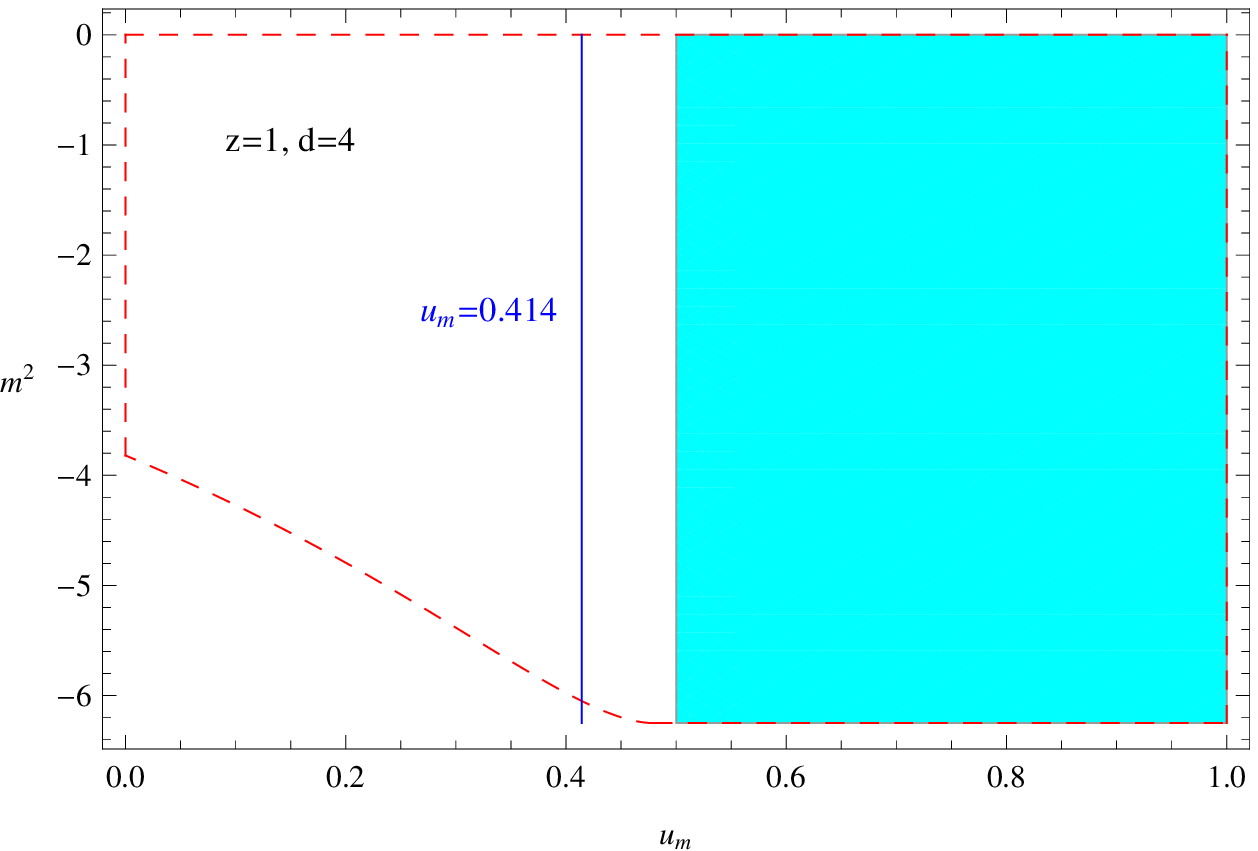}\vspace{0.0cm}
\includegraphics[scale=0.41]{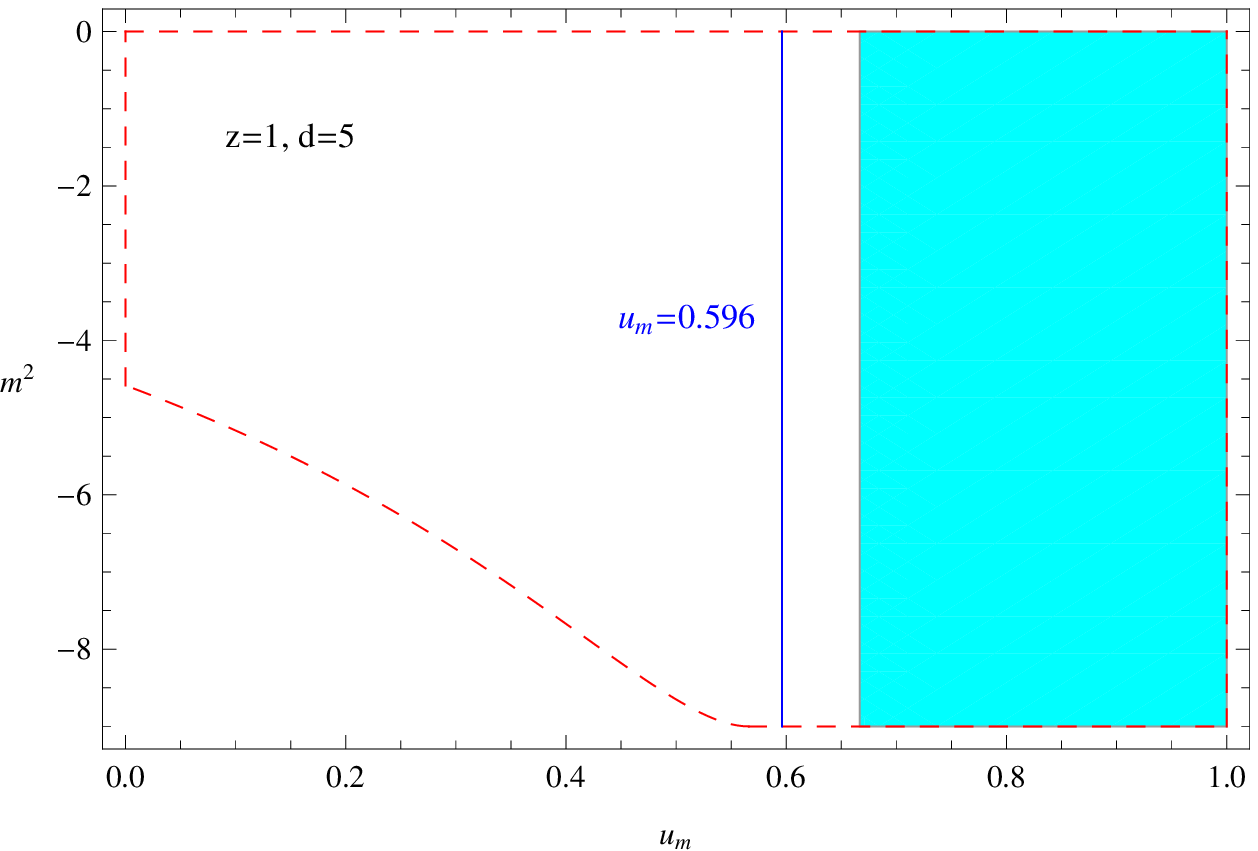}\vspace{0.0cm}
\includegraphics[scale=0.41]{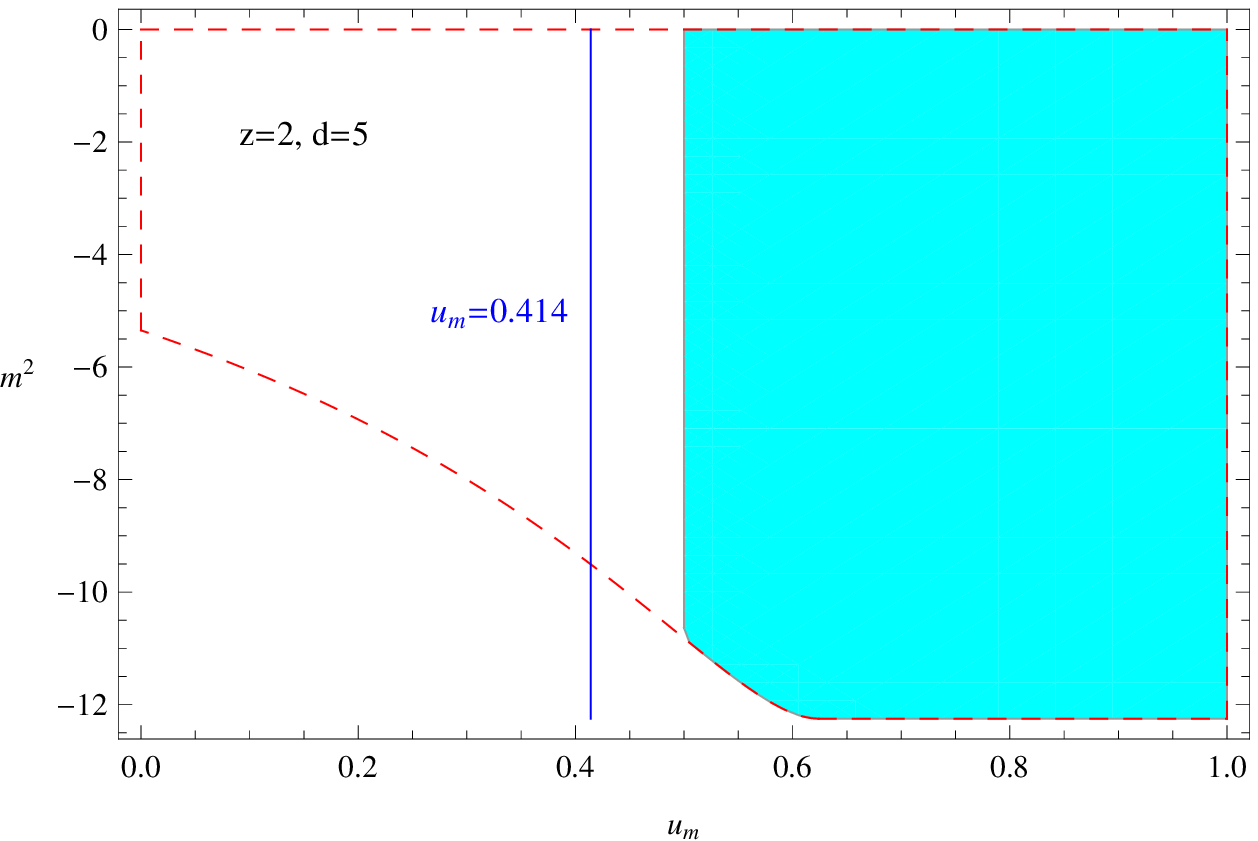}\\ \vspace{0.0cm}
\caption{\label{RegionB} (color online) The allowable range of $u_m$
for different masses if $z\neq d-1$ and $z\neq d-2$. In each panel,
the region surrounded by the red and dashed line is determined by
Eq. (\ref{Range}), and the cyan region corresponds to Eq.
(\ref{RangeTc}). The blue line in each panel represents the fixed
$u_{m}$ given by Eq. (\ref{RelationBc}) and is not in the cyan
region, which indicates that the matching method is invalid in this
case.}
\end{centering}
\end{figure}

\begin{figure}[H]
\begin{centering}
\includegraphics[scale=0.41]{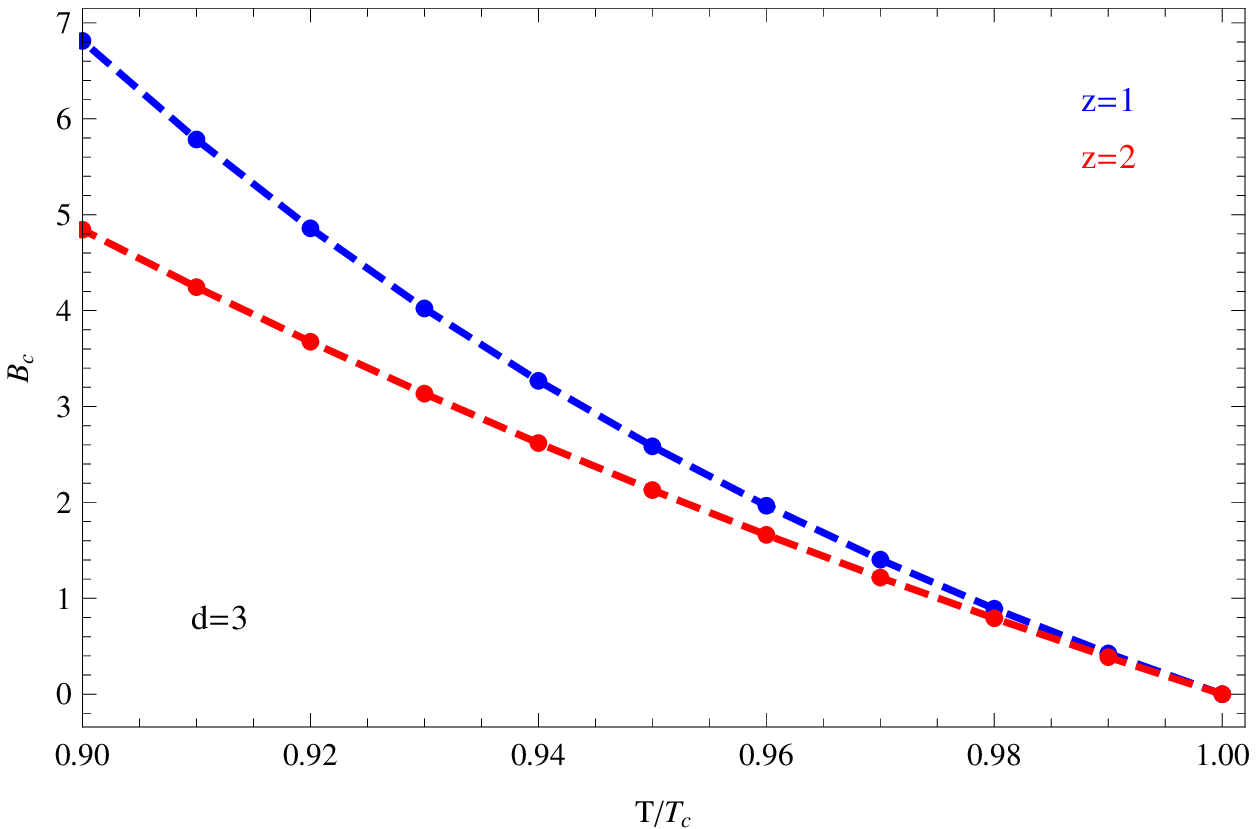}\vspace{0.0cm}
\includegraphics[scale=0.41]{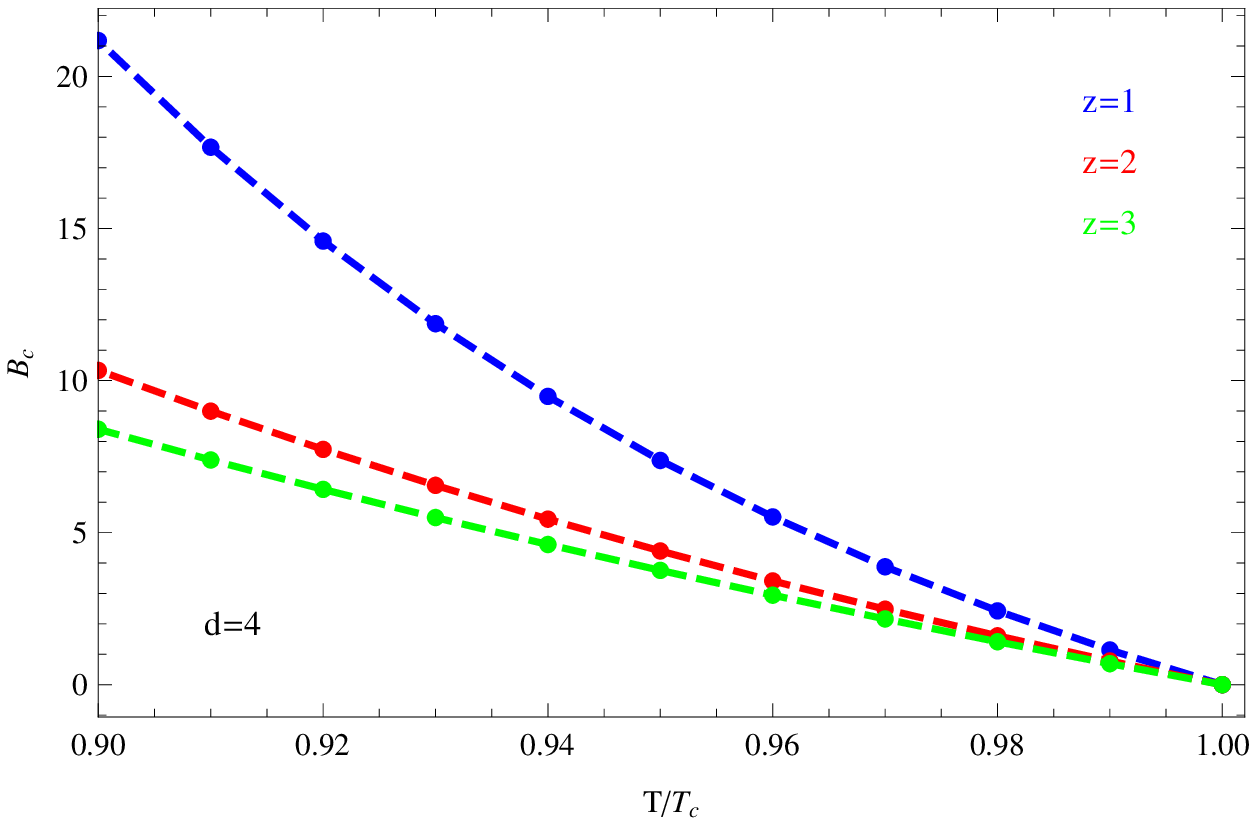}\vspace{0.0cm}
\includegraphics[scale=0.41]{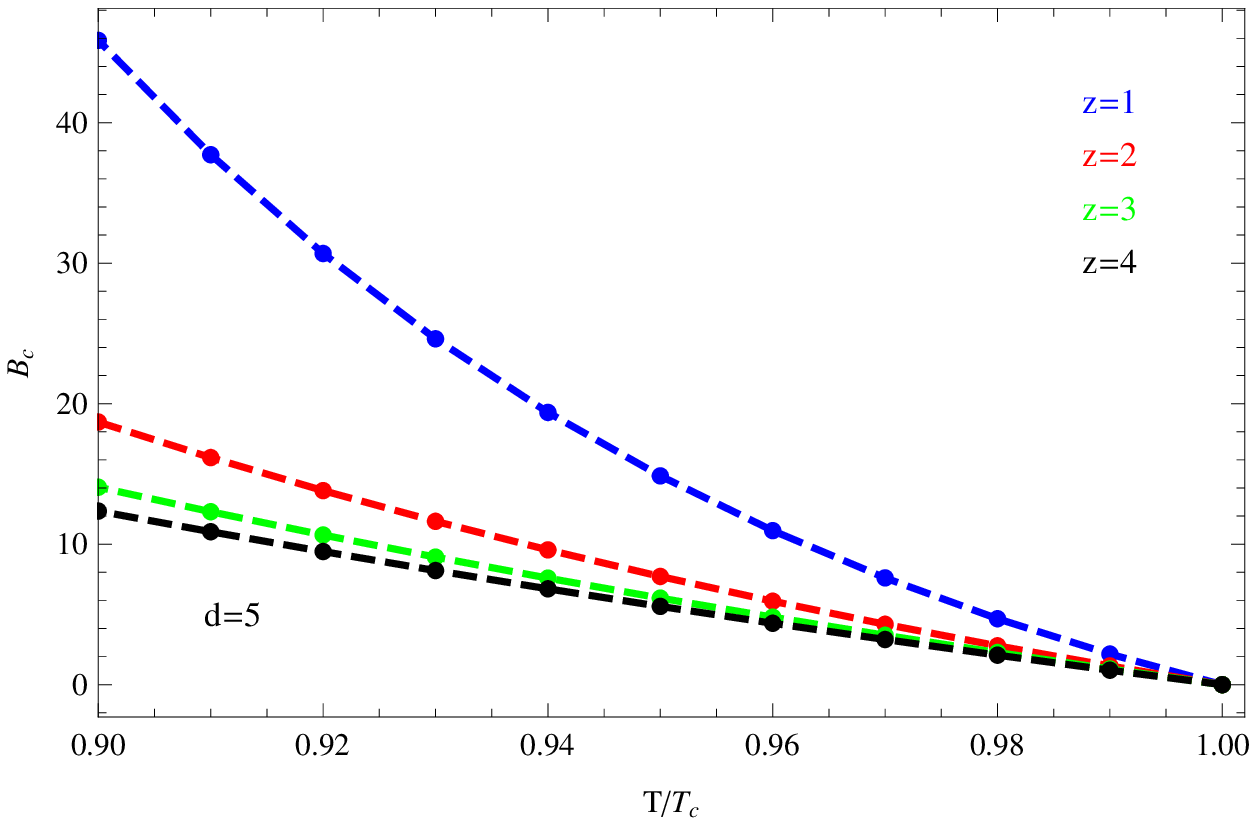}\\ \vspace{0.0cm}
\caption{\label{NumericalBc} (color online) The critical magnetic
field as a function of $T/T_c$ obtained by using the numerical
shooting method. We choose the mass of the scalar field by
$m^{2}=-3$ for the fixed $d=3$, $d=4$ and $d=5$ and set $r_+=1$ in
the numerical computation. The lines from top to bottom are for
$z=1$ (blue), $z=2$ (red), $z=3$ (green) and $z=4$ (black),
respectively.}
\end{centering}
\end{figure}

However, when the constraint (\ref{RelaZD}) can not be satisfied,
for example, $z=1$ and $d=4$, if and only if $u_m=0.414$, we have
$B_{c}=0$ at $T=T_{c}$, which is reminiscent of that seen for the
holographic superconductor in Gauss-Bonnet gravity with Born-Infeld
electrodynamics \cite{Cui}. But from Fig. \ref{RegionB}, this fixed
matching point $u_m$ is not in the allowable region for the correct
critical temperature $T_{c}$, which implies that we can not obtain
the correct expression of the critical magnetic field $B_{c}$.
Obviously, we can observe the same phenomenon for the case $z=1$ and
$d=5$, $z=2$ and $d=5$ in Fig. \ref{RegionB} and other cases when
the constraint (\ref{RelaZD}) can not be satisfied, i.e., $z\neq
d-1$ and $z\neq d-2$. However, we find that the critical magnetic
field $B_{c}$ decreases as $T/T_{c}$ goes up and vanishes at
$T=T_{c}$ from Fig. \ref{NumericalBc} where we use the numerical
shooting method to solve Eq. (\ref{PsiB}) and obtain the critical
magnetic field for different $z$ with the fixed $d=3$, $d=4$ and
$d=5$. Thus, we argue that the matching method is not always valid
to explore the effect of the external magnetic field on the
holographic superconductor with Lifshitz scaling, for example,
$z\neq d-1$ and $z\neq d-2$. In physics, this implies that we can
not ensure the Ginzburg-Landau relation and the correctness of
physical solutions $\psi(u)$ and $\phi(u)$ simultaneously in our
analytic treatment for these cases.

\begin{figure}[H]
\begin{centering}
\includegraphics[scale=0.7]{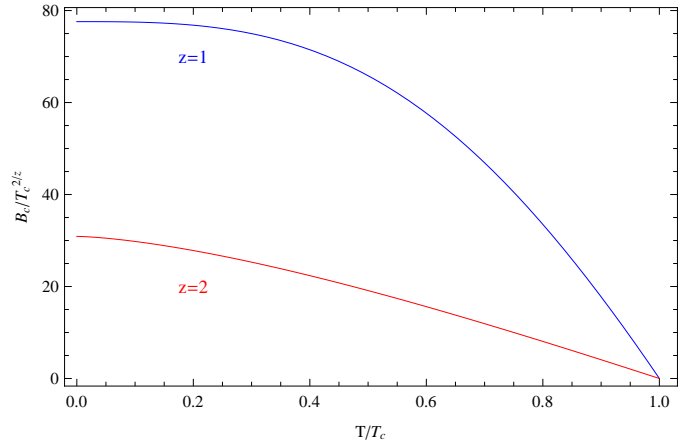}\hspace{0.2cm}%
\caption{\label{Bc} (color online) The critical magnetic field as a
function of $T/T_c$ obtained by using the analytic matching method.
We choose the mass of the scalar field by $m^{2}=-3$ for the fixed
$d=3$ and $u_m=1/2$. The top line corresponds to $z=1$ (blue) and
bottom one is $z=2$ (red).}
\end{centering}
\end{figure}

Using the condition (\ref{RelationBc}), we can obtain the critical
magnetic field of the superconductor with Lifshitz scaling
\begin{eqnarray}
B_{c}\simeq\left(\frac{4\pi
T_{c}}{z+d}\right)^{\frac{2}{z}}\frac{1}{(u_m-1) [(\Delta -2)
u_m-\Delta ]}\left[\sqrt{(\beta^{2}-\gamma)
+\gamma\left(\frac{T}{T_{c}}\right)^{\frac{2d}{z}}}-\beta\left(\frac{T}{T_{c}}\right)^{\frac{d}{z}}\right].
\label{BcFinalA}
\end{eqnarray}
As an example, we choose $z=1$ and $z=2$ with $d=3$, $m^2=-3$ and
$u_m=1/2$ which satisfies the range (\ref{RangeTc}) and constraint
(\ref{RelaZD}), and then have
\begin{equation}
B_c(z=1)\simeq\frac{1}{5}\pi ^2{T_c}^2
\left[\sqrt{1545+856\left(\frac{T}{T_c}\right)^6}-49\left(\frac{T}{T_c}\right)^3\right],
\end{equation}
\begin{equation}
B_c(z=2)\simeq\frac{9-\sqrt{13}}{85} \pi
T_c\left[\sqrt{2}\sqrt{6053+1651\sqrt{13}+10(221+27\sqrt{13})\left(\frac{T}{T_c}\right)^3}-(113+17\sqrt{13})\left(\frac{T}{T_c}\right)^{\frac{3}{2}}\right].
\end{equation}
In Fig. \ref{Bc}, we present the critical magnetic field $B_{c}$ as
a function of $T/T_c$ for different dynamical exponent $z$. It is
clearly shown that the critical magnetic field $B_c$ decreases as we
amplify $z$, which is qualitatively in good agreement with the
numerical results shown in Fig. \ref{NumericalBc}. This suggests
that the dynamical exponent $z$ does have effects on the critical
magnetic field. It should be noted that our result (\ref{BcFinalA})
can reduce to the case discussed in Ref. \cite{Ge} when we take
$z=1,~d=2~,m^2=-2$ and $u_m=1/2$, i.e., $B_c\simeq
\frac{16\pi^2}{9}T_c^2\left[\sqrt{7}\sqrt{4+3\left(\frac{T}{T_c}\right)^4}-7\left(\frac{T}{T_c}\right)^2\right]$.

With the range (\ref{RangeTc}) and constraint (\ref{RelaZD}), when
$T \thicksim T_c$ we can have $B_c\varpropto (1-T/T_c)$ for
different Lifshitz scaling, spacetime dimension and scalar mass,
which agrees well with the Ginzburg-Landau theory. Note that the
relation is independent of $z$, which indicates that the dynamical
exponent $z$ can not modify it. The result may be natural since we
are working in the large $N$ limit. Thus, we can conclude that, for
the case $1\leqslant z<d$, the Ginzburg-Landau theory still holds in
Lifshitz black hole.

\section{Conclusions}

As a very good technique, the matching method can provide us an
analytic understanding of the holographic superconductors in a
straightforward way and help to confirm the numerical result. In
this work, we have used the matching method to investigate the
holographic superconductors with Lifshitz scaling and discussed the
effectiveness of this analytic method. For the cases $1\leqslant
z<d$ considered here, we found that the critical temperature
decreases with the increase of the dynamical exponent $z$, which
shows that Lifshitz scaling makes the condensation harder to occur.
Our analytic result can be used to back up the numerical
computations in the holographic superconductors with Lifshitz
scaling. Furthermore, we analytically studied the holographic
superconductor with Lifshitz scaling in an external magnetic field.
It is interesting to note that, for the case of $z=d-1$ and $z=d-2$,
in order to avoid a breakdown of the matching method we have to
choose the matching point in an appropriate range which depends on
Lifshitz scaling, spacetime dimension and scalar mass. We argued
that the physical conditions lead to the matching range in the
analytic treatment. In this case we observed that a larger $z$
results in a smaller upper critical magnetic field, which is
consistent with the numerical results. This shows that the dynamical
exponent $z$ does have effects on the upper critical magnetic field.
We also reproduced the well-known relation $B_{c}\propto(1-T/T_c)$
from the Ginzburg-Landau theory even in Lifshitz black hole, which
shows that the Lifshitz scaling can not modify this relation. The
result may be natural since we are working in the large $N$ limit.
However, for other cases, i.e., $z\neq d-1$ and $z\neq d-2$, the
matching method can not ensure the Ginzburg-Landau relation and the
correctness of physical solutions $\psi(u)$ and $\phi(u)$
simultaneously, and fails to give the correct expression of the
critical magnetic field of the holographic superconductor with
Lifshitz scaling. The fact implies that the matching method is not
always powerful to explore the effect of the external magnetic field
on the holographic superconductors. The extension of this work to
the fully backreacted spacetime would be interesting. But since the
backreacted solutions are usually not easy to master and the
restricted conditions for the matching method should be
reconsidered, we will leave it for further study.

\begin{acknowledgments}

This work was supported by the National Natural Science Foundation
of China under Grant Nos. 11275066 and 11175065; the National Basic
Research of China under Grant No. 2010CB833004; PCSIRT under Grant
No. IRT0964; Hunan Provincial Natural Science Foundation of China
under Grant Nos. 12JJ4007 and 11JJ7001; and Hunan Provincial
Innovation Foundation For Postgraduate under Grant No. CX2014A009.
Qiyuan Pan thanks the Kavli Institute for Theoretical Physics China
for hospitality in the revised stages of this work.

\end{acknowledgments}

\end{document}